\documentclass[12pt]{article}
\usepackage{epsfig}
\usepackage{amsfonts}
\usepackage{amssymb}
\title{Note on clock synchronization and Edwards transformations } 
\author{{\bf Piotr Kosi\'nski}\thanks{supported by {\L}\'od\'z.University grant no. 795 }\\  
  \it Department of Theoretical Physics II \\
\it University of {\L}\'od\'z \\
\it Pomorska 149/153, 90 - 236 {\L}\'od\'z, Poland.\\
\it e-mail: pkosinsk@uni.lodz.pl}
\date{}

\begin{document}
\maketitle

Edwards transformations relating inertial frames with arbitrary clock synchronization are reviewed and put in more general 
setting. Their group-theoretical context is described.

\vspace{24pt}

Key words: clock synchronization, Edwards transformations, Lorentz group, nonlinear realizations

\newpage
\section{Introduction}
There has been a long-standing discussion concerning foundations of special relativity (SR) (see, for example, Refs. \cite{
b1} $\div$\ \cite{b3} ). As a result  at present most consider that the basic assumptions of SR can be formulated in a 
way independent on any convention concerning clock synchronization. One can also hardly believe that Einstein was unaware 
of this fact, although he used the specific, simple and elegant, synchronization scheme. Within this scheme the notion of 
simultaneity has relative, i.e. depending on the reference frame, character. The absence of absolute simultaneity is slightly 
disturbing if confronted with causality principle because the latter is formulated in terms of time ordering for 
space-time events. 
Not only seems the causality principle to be observer-dependent but also synchronization (i.e. convention) -dependent which 
is even more serious. The way out of this dilemma is simple and well-known: once we assume that all particles and interactions 
propagate inside (or on the surface of) the light cone the only relevant structure entering the theory is the geometry of 
the set of light cones including the possibility of invariant distinction between "past" and "future" cones which allows 
for invariant definition of causality. No synchronization scheme, always based on some convention, is needed; in particular, 
we don't have to refer to the notion of simultaneity. This is because the light cones are geometric objects not depending on the 
choice of coordinates in space-time. The theory becomes elegant, simple and explicitly convention-independent. 

There are sometimes claims that the situation changes when quantum theory enters the game and some synchronization is then 
distinguished. We don't believe this is the case but we will not dwell on this problem here. 

Once the propagation outside the light cones is allowed we are faced with serious troubles. Such exotic signals could be 
used to synchronize the distant clocks and the problem whether the choice of particular synchronization scheme is a matter 
of convention becomes more complicated. The clarity and elegance of Einstein's SR is lost and the theory lacks nice 
geometric interpretation.

All these conclusion, scattered in the literature were clearly expressed more than forty years ago in the elegant and 
concise paper due to Edwards \cite{b4}. Since that time a number of papers have appeared (see, for example \cite{b5}, 
\cite{b6}) where some partial results of Ref. \cite{b4} are discussed. The aim of the present paper is to review the results 
of \cite{b4} by putting them in more general setting and explaining their group-theoretical meaning.
\section{Synchronization}
Let us consider some inertial reference frame which includes also a definite synchronization of distant clocks.
 We shall assume that 
the space is homogeneous but anisotropic in the sense that the light velocity depends only on the direction of ray 
propagation. 
Let $c(\vec{n})$\ be the light velocity in the direction $\vec{n}\;(\mid \vec n\mid =1)$. Our basic assumption is that,
in the inertial frame, the 
average speed over any closed path equals always $c$, the velocity of light in Einstein's theory. This assumption is 
obviously independent on the choice of synchronization.\\

First, we rederive the Edwards result concerning the form of the function $c(\vec n)$. Consider any piecewise smooth closed 
oriented path $\gamma $; in principle, we should rather consider piecewise linear paths realized by the set of properly 
placed mirrors but this is irrelevant. \\

Our assumption implies the following equality to hold for any such path
\begin{eqnarray}
\oint\limits_{\gamma }\frac{ds}{c(\vec n)}=\frac{1}{c}\oint\limits_{\gamma }ds; \label{w1}
\end{eqnarray}
here $\vec n$\ points in the tangent direction at a given point of the curve $\gamma $. \\
Introducing the dimensionless function
\begin{eqnarray}
R(\vec n)\equiv 1-\frac{c}{c(\vec n)} \label{w2}
\end{eqnarray}
one rewrites (\ref{w1}) in the form
\begin{eqnarray}
\oint\limits_{\gamma }R(\vec n)ds=0 \label{w3}
\end{eqnarray}
which holds for any closed path $\gamma $. Eq.(\ref{w3}) implies, as usual, that for any path $\gamma $\ starting at $P_0$\ 
and terminating at $P_1$\ the integral
\begin{eqnarray}
\int\limits_{\gamma }R(\vec n)ds \label{w4}
\end{eqnarray}
depends only on the initial and final points $P_0$\ and $P_1$. Fixing $P_0$\ one defines
\begin{eqnarray}
F(\vec x)=\int\limits_{P_0}^{\vec x}R(\vec n)ds \label{w5}
\end{eqnarray}
where the integral is taken along arbitrary path connecting $P_0$\ and $\vec x$. Eq.(\ref{w5}) implies, in turn, the following 
equality to hold for any path $\gamma $\
\begin{eqnarray}
\int\limits_{\gamma }R(\vec n)ds=\int\limits_{\gamma }\vec \nabla F(\vec x)\cdot \vec nds \label{w6}
\end{eqnarray}
Therefore,
\begin{eqnarray}
R(\vec n)=\vec \nabla F(\vec x)\cdot \vec n  \label{w7}
\end{eqnarray}
which is only possible provided
\begin{eqnarray}
R(\vec n)=\vec k\cdot \vec n, \;\;\;  F(\vec x)=\vec k\cdot \vec x \;\;\; (+const), \label{w8}
\end{eqnarray}
$\vec k$\ being dimensionless and constant. Taking into account the definition of $R(\vec n)$\ one obtains finally
\begin{eqnarray}
c(\vec n)=\frac{c}{1-\vec k\cdot \vec n} \label{w9}
\end{eqnarray}
Following Edwards we assume $c(\vec n)>0$\ (but exclude the case $c(\vec n)=\infty $\ 
for some $\vec n$, considered also by Edwards); consequently
\begin{eqnarray}
\mid \vec k\mid <1  \label{w10}
\end{eqnarray}

Now, it is easy to show that one can always change the synchronization such that the one-way velocity of light is $c$, i. 
$e. \;\; c_e(\vec n)\equiv c$, where $c_e(\vec n)$\ corresponds to new synchronization. To this end define new space-time 
coordinates 
\begin{eqnarray}
&& \vec x_e=\vec x   \nonumber \\
&& t_e=t+\frac{\vec k\cdot \vec x}{c} \label{w11}
\end{eqnarray}
Consider any path $\gamma $\ starting at $\vec x_0$\ and terminating at $\vec x_1$. One has
\begin{eqnarray}
&& \int\limits_{\gamma }\frac{ds}{c}=\int\limits_{\gamma }\frac{R(\vec n)ds}{c}+\int\limits_{\gamma }\frac{ds}{c(\vec n)}\int\limits_{\gamma }\frac{\vec k\cdot d\vec s}{c}+t_1-t_0=  \nonumber \\
&& =\left(t_1+\frac{\vec k\cdot \vec x_1}{c}\right)-\left(t_0+\frac{\vec k\cdot \vec x_0}{c}\right)=t_{1e}-t_{0e} 
 \label{w12}
\end{eqnarray}
One concludes that the assumption that the average light velocity over closed paths equals $c$\ implies the existence of 
Einstein clock synchronization in which one-way velocity of light also always equals $c$. Therefore, traditional special 
relativity 
follows as a special case from the invariance of average closed-path light velocity.
\section{Generalized Lorentz (Edwards) transformations}
It is now straightforward to derive the Edwards transformations relating reference frames with arbitrary synchronization. 
This amounts only to rewrite the standard Lorentz transformations in terms of new variables. First, using eqs.(\ref{w11}) 
we find the relation between velocities in both synchronizations
\begin{eqnarray}
\vec v=\frac{\vec v_e}{1-\frac{\vec k\cdot \vec v_e}{c}},\;\;\;\; \vec v_e=\frac{\vec v}{1+\frac{\vec k\cdot \vec v}{c}} 
 \label{w13}
\end{eqnarray}
Then, from the well-known form of Lorentz transformations \cite{b7} one obtains
\begin{eqnarray}
&& \vec x'=\vec x-\left(\frac{\vec x\cdot \vec v}{\vec v^2}\right)\vec v+\frac{\left((\frac{\vec x\cdot \vec v}{\vec v^2})
(1+\frac{\vec k
\cdot \vec v}{c})-(t+\frac{\vec k\cdot \vec x}{c})\right)\vec v}{\sqrt{(1+\frac{\vec k\cdot \vec v}{c})^2-
\frac{\vec v^2}{c^2}}}  \nonumber \\
&& t'=\left(\frac{\vec k'\cdot \vec v}{c}\right)\left(\frac{\vec x\cdot \vec v}{\vec v^2}\right)-\left(\frac{\vec k'\cdot 
\vec x}{c}\right)+   \label{w14} \\
&& +\frac{\left(1+\frac{(\vec k+\vec k')\cdot \vec v}{c}\right)t+\left(1+\frac{\vec k\cdot  \vec v}{c}\right)
\left(\frac{\vec x\cdot  
\vec k}{c}
-\left(\frac{\vec k'\cdot  \vec v}{c}\right)\left(\frac{\vec x\cdot  \vec v}{\vec v^2}\right)+\left(\frac{\vec k'\cdot  
\vec v}{c}\right)\left(\frac{\vec k \cdot \vec x}{c}\right)-\left(\frac{\vec x\cdot  \vec v}{c^2}\right)\right)}
{\sqrt{(1+\frac
{\vec k\cdot \vec v}{c})^2-\frac{\vec v^2}{c^2}}}  \nonumber
\end{eqnarray}
here $\vec v$\ is the relative velocity between frames as measured in the first frame while $\vec k$\ and $\vec k'$\ 
define synchronizations in both frames.
 
As mentioned 
above, one of the main features of Einstein synchronization is the relative character of simultaneity. Now, one can pose the 
question if there exists synchronization procedure making the notion of simultaneity absolute. To answer it we put $t=0$\ 
and ask under what condition this implies $t'=0$. It follows from eqs.(\ref{w14}) that this is possible only provided
\begin{eqnarray}
 && 0= 
 \frac{\left(1+\frac{\vec k\cdot \vec v}{c}\right)\frac{\vec k}{c}-\frac{\vec v}{c^2}+\left(\frac{\vec k'\cdot \vec v}
{c}\right)\frac{\vec k}{c}-\left(1+\frac{\vec k\cdot \vec v}{c}\right)\left(\frac{\vec k'\cdot \vec v}{c}\right)\frac{\vec v}
{\vec v^2}}{\sqrt{(1+\frac{\vec k\cdot \vec v}{c})^2-\frac{\vec v^2}{c^2}}}+ \nonumber \\
&& +\left(\frac{\vec k'\cdot \vec v}{c}\right)\frac
{\vec v}{\vec v^2}-\frac{\vec k'}{c} \label{w15}
\end{eqnarray}
Solving (\ref{w15}) for $\vec k'$\ one obtains
\begin{eqnarray}
&& \frac{\vec k'}{c}= 
\left(\frac{\vec k\cdot \vec v}{c}\right)\left(1+\frac{\vec k\cdot \vec v}{c}\right)\frac{\vec v}{\vec v^2}
-\frac{\vec v}{c^2}+ \nonumber \\
&& +\sqrt{(1+\frac{\vec k\cdot \vec v}{c})^2-\frac{\vec v^2}{c^2}}\left(\frac{\vec k}{c}-(\frac{\vec k\cdot 
\vec v}{c})\frac{\vec v}{\vec v^2}\right) \label{w16}
\end{eqnarray}
In order to put eq.(\ref{w16}) in more familiar form we define new velocity variable $\vec u\equiv k\vec c\;\;(\mid \vec u
\mid <c)$\ and express $\vec v$\ in terms of $\vec v_e$\ (of. eq.(\ref{w13})). Then (\ref{w16}) takes the form
\begin{eqnarray}
\vec u'=\frac{\sqrt{1-\frac{\vec v^2_e}{c^2}}\vec u-\vec v_e+\left(1-\sqrt{1-\frac{\vec v^2}{c^2}}\right)\left(\frac
{\vec u\cdot \vec v_e}{\vec v^2_e}\right)\vec v_e}{1-\frac{\vec u\cdot \vec v_e}{c^2}}  \label{w17}
\end{eqnarray}
which is just the Einstein addition formula for velocities \cite{b7}. Therefore, we can think on $\vec u$\ as the velocity of
 the actual reference frame with respect to some fixed frame called the "preferred" one; note that $\vec u=\vec u_e$\ 
because in the preferred frame synchronization reduces to the standard one. \\

One concludes that the most general definition of absolute simultaneity is that one selects a fixed but arbitrary reference 
frame ("preferred" frame) with standard synchronization and calls two events simultaneous if they are simultaneous in 
preferred frame.
\section{Group theory of Edwards transformations}
Let us remind some facts about the geometry of Lorentz group. Denoting $x^0_e\equiv ct_e$\ and $g^{\mu \nu }=g_{\mu \nu }
=diag(+---)$\ one defines Lorentz transformation as a general linear transformation
\begin{eqnarray}
x'^{\mu }_e=\Lambda ^{\mu }_{\;\;\nu }x^{\nu }_e  \label{w18}
\end{eqnarray}
leaving invariant the quadratic form
\begin{eqnarray}
x^2_e\equiv g_{\mu \nu }x^{\mu }_ex^{\nu }_e  \label{w19}
\end{eqnarray}
Eqs. (\ref{w18}), (\ref{w19}) imply the following constraints on $\Lambda ^{\mu }_{\;\;\nu }$:
\begin{eqnarray}
g_{\mu \nu }\Lambda ^{\mu }_{\;\;\alpha }\Lambda ^{\nu }_{\;\;\beta }=g_{\alpha \beta }   \label{w20}
\end{eqnarray}
Therefore, the general Lorentz transformation depends on six parameters which parametrize Lorentz boosts (components of 
relative velocity $\vec v_e$\ ) and rotations (three angles). \\

Any Lorentz matrix can be represented as the product of two matrices representing pure boost and rotation. Indeed, the 
following identity can be easily checked using eqs.(\ref{w20})
\begin{eqnarray}
\Lambda =\tilde \Lambda \cdot \mathcal{R}  \label{w21}
\end{eqnarray}
where
\begin{eqnarray}
&& \tilde \Lambda ^0_{\;\;0}=\Lambda ^0_{\;\;0},
\;\;\;\tilde \Lambda ^0_{\;\;i}=\tilde \Lambda ^i_{\;\;0}=\Lambda ^i_{\;\;0}     \nonumber \\
&& \tilde \Lambda ^i_{\;\;j}=\delta ^i_{\;\;j}+\frac{\Lambda ^i_{\;\;0}\Lambda ^j_{\;\;0}}
{\Lambda +\Lambda ^0_{\;\;0}}  \label{w22}
\end{eqnarray}
while
\begin{eqnarray}
\mathcal{R}=\left[\begin{array}{cc}
1 & 0 \\
0 & R\end{array}\right] \label{w23}
\end{eqnarray}
with
\begin{eqnarray}
R^i_j=\Lambda ^i_{\;\;j}-\frac{\Lambda ^i_{\;\;0}\Lambda ^0_{\;\;j}}{1+\Lambda ^0_{\;\;0}}  \label{w24}
\end{eqnarray}
Both $\tilde \Lambda $\ and $\mathcal{R}$\ obey (\ref{w20}) and $R$\ is an orthogonal matrix. The latter means that $R$\ describes 
rotation. On the other hand, $\tilde \Lambda $\ is symmetric. Due to the relation
\begin{eqnarray}
(\tilde \Lambda ^0_{\;\;0})^2-\sum\limits_{i=1}^3(\tilde \Lambda ^i_{\;\;0})^2=1  \label{w25}
\end{eqnarray}
one can choose the parameterisation
\begin{eqnarray}
\tilde \Lambda ^0_{\;\;0}=\frac{1}{\sqrt{1-\frac{\vec v^2_e}{c^2}}},\;\;\; 
\tilde \Lambda ^i_{\;\;0}=\frac{\frac{-v^i_e}{c}}{\sqrt{1-\frac{\vec v^2_e}{c^2}}},\;\;\;
\mid \vec v_e\mid <c   \label{w26}
\end{eqnarray}
which leads to standard Lorentz boost (eq.(\ref{w14}) with $\vec k=0=\vec k'$.\\
Let $\Lambda (\vec v_e)$\ be pure boost; $\Lambda (\vec v_e)$\ is a symmetric matrix. For generic $\vec v_{e1},\;\vec v_
{e2},\;\Lambda (\vec v_{e1})$\ and $\Lambda (\vec v_{e2})$\ do not commute. Therefore, their product is, in general, not 
symmetric, i.e. it is not a pure boost; in fact, using the decomposition (\ref{w21}) one finds
\begin{eqnarray}
\Lambda (\vec v_{e1})\Lambda (\vec v_{e2})=\Lambda (\vec v_{e1}\oplus \vec v_{e2})\mathcal{R}(\vec v_{e1},\vec v_{e2}) 
\label{w27}
\end{eqnarray}
here $\vec v_{e1}\oplus \vec v_{e2}$\ denotes Einstein sum of velocities while $\mathcal{R}(\vec v_{e1},\vec v_{e2})$\ 
describes the 
rotation giving rise to the so called Thomas precession.\\
It is also easy to check that
\begin{eqnarray}
\mathcal{R}\Lambda (\vec v_e)=\Lambda (R\vec v_e)\mathcal{R} \label{w28}
\end{eqnarray}
The general composition rule can be now obtained from (\ref{w21}), (\ref{w27}), (\ref{w28}):
\begin{eqnarray}
&& \Lambda _1\Lambda _2=(\tilde \Lambda (\vec v_{e1})\mathcal{R}_1)(\tilde \Lambda (\vec v_{e2})\mathcal{R}_2)  \nonumber \\
&& =\tilde \Lambda (\vec v_{e1})\tilde \Lambda (R_1\vec v_{e2})\mathcal{R}_1\mathcal{R}_2   \label{w29}  \\
&& =\tilde \Lambda (\vec v_{e1}\oplus R_1\vec v_{e2})\mathcal{R}(\vec v_{e1},R_1\vec v_{e2})\mathcal{R}_1\mathcal{R}_2 \nonumber
\end{eqnarray}

Let us now explain the group-theoretical meaning of Edwards transformations (\ref{w14}). Physically it is obvious that they 
are equivalent to the standard Lorentz ones. However, they cannot be obtained from the latter by a simple redefinition 
of space-time coordinates. In fact, such redefinition 
involves the quantity $\vec k$\ which, in turn, depends on the choice of reference frame. So there exists no change of 
space-time coordinates reducing Edwards transformations to Lorentz ones. On the other hand, the physical equivalence of 
synchronization procedures must be reflected somehow in mathematical formalism.\\

We start with the following simple remark. Let us fix some reference frame with standard synchronization. Any other standard 
frame can be obtained by applying the uniquely defined Lorentz transformation. Therefore, the totality of all standard 
inertial reference frames can be parametrized by the coordinates of Lorentz group manifold, i.e. by three components of 
relative velocity $\vec v_e$\ and three angles of the rotation matrix $R$. \\

The totality of all coordinates of space-time points in all inertial frames is now parametrized as follows. Let us select 
an arbitrary reference frame and let $\Lambda (\vec u,\overline{R})=\tilde \Lambda (\vec u)\overline{\mathcal{R}}$\ be the unique Lorentz 
transformation which leads to this frame when applied to the preferred one. The space-time coordinates with respect to the 
preferred frame are denoted by $x^{\mu }_p$. For any space-time point its coordinates with respect to the actual reference 
frame will be parametrized by $\vec u,R$\ and $x^{\mu }_p$\ -the coordinates of this point in preferred frame. It is easy 
to check that the action of the Lorentz group, when expressed in new coordinates, reads
\begin{eqnarray}
&& \;\;\;\;\;\;\;\;\;\;\;\;\;\;\;\;\;\;\;\;\;\vec u\rightarrow \vec v_e\oplus R\vec u  \nonumber \\
&& \Lambda (\vec v_e,R): \;\;\;\;\; \overline{\mathcal{R}}\rightarrow \mathcal{R}(\vec v_e,R\vec u)
\mathcal{R}\overline{\mathcal{R}}  \label{w30} \\
&&  \;\;\;\;\;\;\;\;\;\;\;\;\;\;\;\;\;\;\;\; x^{\mu }_p\rightarrow x^{\mu }_p  \nonumber
\end{eqnarray}
which means that $(\vec u,\overline{R})$\ transform according to the left action of Lorentz group on itself while $x^{\mu }
_p$\ are Lorentz invariants. The reader familiar with the theory of nonlinear group realizations \cite{b8} will immediately 
recognise in (\ref{w30}) the canonical form of nonlinear realizations of Lorentz group which linearize on its trivial 
subgroup. Replacing the coordinates $x^{\mu }_p$\ by 
\begin{eqnarray}
x^{\mu }_e\equiv \Lambda ^{\mu }_{\;\;\nu }(\vec u,\overline{R})x^{\nu }_p  \label{w31}
\end{eqnarray}
one finds that $x^{\mu }_e$\ transform in the standard way under Lorentz transformations, again in accordance with the 
general theory \cite{b8}.\\

The quantities $\vec k$\ defining synchronization depend on the choice of reference frame,
\begin{eqnarray}
\vec k=\vec k(\vec u,\overline{R})  \label{w32}
\end{eqnarray}
Now, according to the formula (\ref{w11}) one defines new coordinates
\begin{eqnarray}
x^{\mu }\equiv T^{\mu }_{\;\;\nu }(\vec u,\overline{R})\Lambda ^{\nu }_{\;\;\alpha }(\vec u,\overline{R})x^{\alpha }_p 
\label{w33}
\end{eqnarray}
where
\begin{eqnarray}
&& T^0_{\;\;0}(\vec u,\overline{R})=1,\;\;\;\; T^0_{\;\;i}(\vec u,\overline{R})=-k^i(\vec u,\overline{R}) \\ \nonumber
&& T^i_{\;\;0}(\vec u,\overline{R})=0, \;\;\;\; T^i_{\;\;j}(\vec u,\overline{R})=\delta ^i_j \label{w34}
\end{eqnarray}
Since $T$\ is invertible, $(\vec u,\overline{R},x^{\mu })$\ provide equally good choice of parametrization as $(\vec u,
\overline{R},x^{\mu }_p)$\ and $(\vec u,\overline{R},x^{\mu }_e)$. The transformation rules involve now the left action 
of Lorentz group for $(\vec u,\overline{R})$\ and Edwards transformations (\ref{w14}) for $x^{\mu }$. Contrary to the 
canonical choice $(x^{\mu }_p)$\ and the one that linearizes the action of the whole group in the spirit of Ref.\cite{b8},  
$(x^{\mu }_e)$, in this case the variables $(\vec u,\overline{R})$\ and $x^{\mu }$\ do not decouple. \\

For some particular 
choices of $\vec k$\ one can obtain more general form of nonlinear realizations of Lorentz group. This is, for example, 
the case for the $\vec k$\ defined as
\begin{eqnarray}
\vec k=\frac{\vec u}{c}  \label{w35}
\end{eqnarray}
(cf. discussion below eq. (\ref{w16})). Then $\vec k$\ is a function on coset space Lorentz/rotation subgroup. One considers 
the totality of reference frames obtained by applying \underline{pure boosts} to the preferred one:
\begin{eqnarray}
x^{\mu }_e=\Lambda ^{\mu }_{\;\;\nu }(\vec u)x^{\nu }_p\equiv \tilde \Lambda ^{\mu }_{\;\;\nu }(\vec u)x^{\nu }_p \label{w36}
\end{eqnarray}
Then
\begin{eqnarray}
&& \Lambda ^{\mu }_{\;\;\nu }(\vec v,R)x^{\nu }_e=\Lambda ^{\mu }_{\;\;\nu }(\vec v,R)\Lambda ^{\nu }_{\;\;\varrho }
(\vec u)x^{\varrho }_p \\  \nonumber
&& =\tilde \Lambda ^{\mu }_{\;\;\nu }(\vec v\oplus R\vec u)\mathcal{R}^{\nu }_{\;\;\varrho }(\vec v,R\vec u)
\mathcal{R}^{\varrho }_{\;\;\sigma }
x^{\sigma }_p \label{w37}
\end{eqnarray}
which implies that $\vec u$\ transform as "goldstonic" and $x^{\mu }_p$\ as adjoint coordinates according to the 
terminology of Ref.\cite{b8}. The action linearizes on rotation subgroup while $x^{\mu }_e$\ are the new coordinates on 
which full Lorentz group acts linearly, again in the spirit of general theory. Let us note that one can even here consider the 
totality of all inertial frames by replacing eq.(\ref{w36}) by eq.(\ref{w31}). The rotation parameters entering $\overline
{R}$\ play, however, the role of spectators: for any $\overline{R}, \; \overline{R}^{\mu }_{\;\;\nu }x^{\nu }_p$\ are adjoint 
coordinates.

Edwards transformations are  again obtained by passing to the new coordinates
\begin{eqnarray}
x^{\mu }=T^{\mu }_{\;\;\nu }(\vec u)\Lambda ^{\nu }_{\;\;\alpha }(\vec u)x^{\alpha }_p \label{w38}
\end{eqnarray}
Let us notice that in Ref.\cite{b6} the Edwards transformations for the very particular choice $\vec k=\frac{\vec u}{c}$\ 
were also considered from the point of view of nonlinear realizations.

\vspace{12pt}

{\large\bf Acknowledgement}

Thanks are due to Serafin Kosi\'nski for criticism concerning the first version of the paper.


\begin{thebibliography}{99}
\bibitem{b1}
H.Reichenbach, The Philosophy of Space and Time, Dover Publ. Inc., N.Y. 1958
\bibitem{b2}
A.Grunbaum, Logical and Philosophical Foundations of the Special Theory of Relativity, in: Philosophy of Sciencies, 
ed. A.Danto, S.Morgenbesser, Meridian Books, N.Y. 1960
\bibitem{b3}
H.Reichenbach, Axiomatization of the Theory of Relativity, Univ.Calif. Press, Berkeley, CA1969
\bibitem{b4}
W.F.Edwards, Am. J. Phys. (1963), 482
\bibitem{b5}
Chang, Phys. Lett. {\bf A70}, (1979),1; J.Phys.{\bf A12}, (1979), L203
\bibitem{b6}
J. Rembielinski, Phys. Lett.{\bf A78} (1979), 33; Int.J. Mod. Phys.{\bf A12}, (1997), 1677
\bibitem{b7}
C.M$\o$ller, The Theory of Relativity, Claredon Press, Oxford 1972
\bibitem{b8}
S.Coleman, J.Wess, B.Zumino, Phys. Rev. 177 (1969), 2239


\end{thebibliography}
\end{document}